\documentclass[reprint,superscriptaddress,amsmath,amssymb,aps,linenumbers,floatfix]{revtex4-1}
\usepackage{graphics}
\usepackage{epsfig}
\usepackage{dcolumn}
\usepackage{color}
\usepackage{bm}

\begin{document}


\title{Magnetic Anisotropy of Single Mn Acceptors in GaAs in an External Magnetic Field}

\author{M. Bozkurt}
\email[E-mail:]{Bozkurt.Murat.2010@gmail.com}
\affiliation{Photonics and Semiconductor Nanophysics, Department of Applied Physics, Eindhoven University\\
of Technology, P. O. Box 513, NL-5600 MB Eindhoven, The Netherlands}
\author{M.R. Mahani}
\affiliation{School of Computer Science, Physics and Mathematics, Linnaeus University,391 82 Kalmar, Sweden}
\author{P. Studer}
\affiliation{London Centre for Nanotechnology, University College London (UCL), London, WC1H 0AH, U.K.}
\affiliation{Department of Electronic and Electrical Engineering, UCL, London, WC1E 7JE, UK}
\author{J.-M. Tang}
\affiliation{Department of Physics, University of New Hampshire, Durham, New Hampshire 03824-3520, U.S.A.}
\author{S.R. Schofield}
\affiliation{London Centre for Nanotechnology, University College London (UCL), London, WC1H 0AH, U.K.}
\affiliation{Department of Physics and Astronomy, UCL, London, WC1E 6BT, UK}
\author{ N.J. Curson}
\affiliation{London Centre for Nanotechnology, University College London (UCL), London, WC1H 0AH, U.K.}
\affiliation{Department of Electronic and Electrical Engineering, UCL, London, WC1E 7JE, UK}
\author{M.E Flatt{\'e} }
\affiliation{Department of Physics and Astronomy, University of Iowa, Iowa City, Iowa 52242-1479, U.S.A.}
\author{A. Yu. Silov}
\affiliation{Photonics and Semiconductor Nanophysics, Department of Applied Physics, Eindhoven University\\
of Technology, P. O. Box 513, NL-5600 MB Eindhoven, The Netherlands}
\author{C.F. Hirjibehedin}
\affiliation{London Centre for Nanotechnology, University College London (UCL), London, WC1H 0AH, U.K.}
\affiliation{Department of Physics and Astronomy, UCL, London, WC1E 6BT, UK}
\affiliation{Department of Chemistry, UCL, London, WC1H 0AJ, UK}
\author{C.M. Canali}
\affiliation{School of Computer Science, Physics and Mathematics, Linnaeus University,391 82 Kalmar, Sweden}
\author{P.M. Koenraad}
\affiliation{Photonics and Semiconductor Nanophysics, Department of Applied Physics, Eindhoven University\\
of Technology, P. O. Box 513, NL-5600 MB Eindhoven, The Netherlands}

\date{\today}

\begin{abstract}
We investigate the effect of an external magnetic field on the physical properties of the acceptor hole states 
associated with single Mn acceptors placed near the (110) surface of GaAs. Cross-sectional scanning tunneling microscopy
images of the acceptor local density of states (LDOS) show that the strongly anisotropic hole wavefunction is not
significantly affected by a magnetic field up to 6 T. These experimental results are supported by theoretical calculations based on a tight-binding model of Mn acceptors in GaAs.
For Mn acceptors on the (110) surface and the subsurfaces immediately underneath, we find that an applied magnetic field
modifies significantly the magnetic anisotropy landscape. However the acceptor hole wavefunction is strongly localized
around the Mn and the LDOS is quite independent of the direction of the Mn magnetic moment.
On the other hand, for Mn acceptors placed on deeper layers below the surface, the acceptor hole wavefunction is more 
delocalized and the corresponding LDOS is much more sensitive on the direction of the Mn magnetic moment.
However the magnetic anisotropy energy for these magnetic impurities is large (up to 15 meV),
and a magnetic field of 10 T can hardly change the landscape and rotate the direction of the Mn magnetic moment away
from its easy axis.
We predict that substantially larger magnetic fields are required to observe a significant field-dependence of the
tunneling current for impurities located several layers below the GaAs surface.
\begin{description}
\item[PACS numbers]
75.50.Pp
\end{description}
\end{abstract}
\pacs{Valid PACS appear here}
                          
\maketitle

\section{\label{sec:level1}INTRODUCTION\protect}
Magnetic semiconductors have attracted strong attention in the last decade because of their potential 
to combine opto-electronic and magnetic properties in spintronic devices. 
The most commonly investigated material as a magnetic semiconductor is GaAs doped 
with transition metal Mn-impurities. Mn acts as an acceptor in GaAs and its 
magnetic properties are mainly determined by the magnetic moment of the half filled d-shell \cite{Soviet1982}. 
In highly Mn doped GaAs, the observed ferromagnetism in GaMnAs has been shown to be hole mediated \cite{PhysRevB.63.195205,Jungwirth_2005}, as a result of exchange coupling between the p-like acceptor holes residing in the valence band and the electrons in the d-shell which we will refer to as the Mn core from now on. On the other hand, for applications in spintronic devices, it is important to investigate methods 
to read, set and manipulate the magnetic orientation of the Mn core, especially at the level of a single Mn impurity. Spectacular results have been achieved with optical polarization and manipulation of low Mn doped GaAs/AlGaAs quantum wells \cite{MyersNMAT} and single Mn doped quantum dots \cite{KudelskiPRL,MariettePRL}. Other important work in the field of single spin reading and manipulation has been done for single nitrogen-vacancy centers in diamond \cite{AwschalomNature}.

In this paper, we investigate low-concentration Mn-doped GaAs. Because Mn has strongly coupled 
magnetic and electric properties, spin manipulation by electric fields has been suggested as a possibility in addition to manipulation by magnetic and optical fields. Cross-sectional scanning tunneling microscopy (X-STM) 
has been used in the past to study the Mn acceptor wave function at the atomic scale 
and to manipulate its charge state. The experimental study of the Mn acceptor wavefunction 
by X-STM showed a strongly anisotropic shape of the acceptor wavefunction \cite{yakunin_prl04} as was predicted by tight binding calculations \cite{FlattePRL2004}. 
These experimental and theoretical results  proved that the observed anisotropy 
of the acceptor wavefunction is due to the cubic symmetry of the GaAs crystal.
Additional studies showed that the anisotropy of the Mn acceptor 
wavefunction is also influenced by (local) strain due to a nearby InAs 
quantum dot \cite{paul_nature_2007} or the relaxation of the surface \cite{celebi_prl10}. 

These results indicate that STM can also be an excellent tool to investigate the effects of 
the orientation of the magnetic moment of the Mn core on the 
acceptor wavefunction. In fact, theoretical work\cite{tangflatte_prb05},\cite{scm_MnGaAs_paper1_prb09}
has predicted that the local density of states (LDOS) of the acceptor-hole wavefunction can depend strongly
on the direction of the Mn moment. Since the LDOS is directly related to the tunneling current, these
predictions suggest that it might be possible to control the STM electric current by manipulating the 
individual Mn core spin, for example with an external magnetic field. An X-STM and X-STS study of the energetic level of Mn close to the GaAs [110] cleavage surface has already shown that the 3-fold degeneracy of the J=1 ground level is split because of the reduced symmetry \cite{GarleffPRB2010}.
Magnetic-field manipulation and control of atomic
spins is presently undergoing fast progress, 
showing great promise to selectively address individual atoms \cite{yacoby_nat_phys_2011}. 
Control of atomic spin, combined with the aforementioned sensitivity of the STM current on the dopant magnetic
moment direction, could be a crucial step in realizing multifunctional spin-electronic devices based on individual
atoms. Apart from addressing electrical properties of single magnetic dopants, STM has been shown to be also well capable of positioning individual dopants within a semiconductor surface \cite{yazdani_nat06,Gupta_NanoLett}.

In this paper we will use STM to explore the effect of an external magnetic field on the 
magnetic orientation of the magnetic moment of a single Mn impurity in dilute Mn doped GaAs and compare 
the results with tight-binding model calculations. 
In Section~\ref{review} we present a review of the theoretical work that 
has been published in 2 papers \cite{tangflatte_prb05},\cite{scm_MnGaAs_paper1_prb09}. 
These calculations are based on a tight binding model and show that a change 
in the spin orientation of the Mn core can sometimes give rise to a detectable
change in LDOS of the Mn acceptor wavefunction. 
In Section~\ref{exp_results}  we present experimental results of STM measurements on single Mn impurities in GaAs in a 
magnetic field. We will show that
the LDOS of the Mn acceptor wavefunction is not significantly modified by magnetic fields up to 6 Tesla.
In Sections~\ref{theo_model} and \ref{theo_results} we present theoretical results 
of tight binding modelling of Mn in GaAs 
where a magnetic field has been explicitly included in the Hamiltonian. 
These calculations support our experimental observations and show that 
a dependence of LDOS on external magnetic is in fact expected only for Mn acceptors placed several
layers below the GaAs (110) and can be detected only with stronger magnetic 
fields than the ones presently available.

\section{Review}
\label{review}
Tang et al. \cite{tangflatte_prb05} and Strandberg et al. \cite{scm_MnGaAs_paper1_prb09} 
have reported results of calculations of the dependence the Mn acceptor hole wavefunction 
on the orientations of the Mn magnetic moment. 
The paper by Tang et al. \cite{tangflatte_prb05} describes the Mn LDOS in bulk GaAs 
with an sp$^{3}$ tight binding model in which the Mn core spin is taken in calculation 
by a spin dependent term in the potential at the four nearest neighbor sites in a zinc-blende crystal. 
It is found that the energy spectrum of the Mn is independent of the Mn core spin orientation. 
However, the LDOS of the Mn is found to be depending on the Mn spin orientation. 
A qualitative description of this dependence is given in terms of spin-orbit coupling between the spin of 
the Mn core and the orbital character of the Mn acceptor hole. 
In absence of spin-orbit interaction, the LDOS of the Mn acceptor state would have 
the same T$_{d}$ symmetry as the surrounding zinc-blende crystal. 
However, the spin-orbit coupling is taken into account and the symmetry of 
the Mn acceptor wavefunction is reduced. The contour surface of the acceptor 
LDOS for various Mn core spin directions show that in general, the LDOS has an 
oblate shape with the short axis aligned with the Mn core spin axis. 
For a quantitative comparison with X-STM experiments, cross sectional views of the 
LDOS are calculated in the (110) plane. The largest variation in the cross sectional 
images of the LDOS is seen when the Mn core spin direction changes from [001] or [1$\bar{1}$0] to [110]. 
A variation in LDOS of up to 90$\%$ is predicted by these tight binding calculations 
when the Mn core spin switches from parallel to perpendicular to the (110) surface. 
There is also a small difference of 15$\%$ in the LDOS when the Mn core spin is aligned 
in the two directions parallel to the (110) plane. When the spin of Mn core can be 
changed with an external magnetic field and possibly 
with ESR techniques \cite{schneider_prb87, yacoby_nat_phys_2011}, 
the differences in the LDOS are expected to be visible in an X-STM experiment.

\begin{figure}[h]
\centering
\includegraphics[scale=0.7]{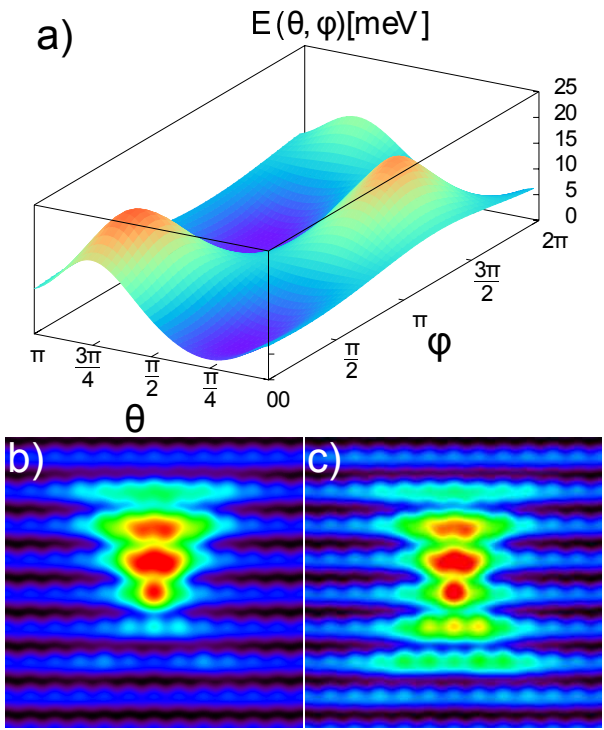}
\caption{(a) Magnetic Anisotropy Energy (MAE) of a Mn acceptor according to a tight binding calculation for strained bulk material. The energy level difference between the easy and the hard axis is about 23 meV, based on a uniform strain estimated in Ref. \cite{celebi_prl10}. The angles $\theta$ and $\phi$ have the same definition as in Fig. \ref{fig:crystal}. (b) Mn LDOS at five atomic layers from the Mn position when the core spin is oriented along the (b) easy axis or (c) hard axis.}  
\label{fig:JMing}
\end{figure}

This model gives a good description of Mn in bulk GaAs but the effect of 
the cleavage surface is completely neglected. In fact it has been shown experimentally\cite{celebi_prl10} that 
the wavefunction of a Mn near the (110) cleavage surface can be strongly affected 
by the strain from the surface relaxation. In the same paper, bulk tight binding calculations support the observation of a broken symmetry near the surface. The surface is taken into account by applying a uniform strain to the bulk model by shifting the Ga lattice with respect to the As lattice. The calculation results presented in that paper are the average of different Mn core spin orientations. In Fig. \ref{fig:JMing}, the same results are presented but for individual Mn core spin orientations. Fig. \ref{fig:JMing} was unpublished in this form. A clear difference in LDOS can be observed when the Mn core spin changes its orientation from the hard axis to the easy axis.

In the paper by Strandberg et al. \cite{scm_MnGaAs_paper1_prb09} the 
reconstructed surface is taken into account for the calculation of the LDOS dependence 
of the Mn acceptor state on the Mn core spin orientation using a tight binding model 
which includes the exchange interaction, spin-orbit coupling and Coulomb interaction. 
This makes a direct comparison with X-STM experiments more justified and the results 
indeed show the same experimentally observed breaking of the symmetry of the wavefunction 
due to the near presence of the surface. In Ref.~\onlinecite{scm_MnGaAs_paper1_prb09} 
Mn acceptors in bulk GaAs
(neglecting the surface) have also been considered. For Mn in bulk GaAs, the energy level 
of the Mn state calculated for different orientations of the Mn core spin 
shows a small magnetic anisotropy, in contrast to the results of
Ref.~\onlinecite{tangflatte_prb05}, where no magnetic anisotropy was found for Mn in bulk. 
The easy axis in Ref.~\onlinecite{scm_MnGaAs_paper1_prb09} for the Mn core spin is oriented along the [001] direction 
whereas the hard axis is found to be lying in the (001) plane. The energy barrier between 
the hard axis and the easy axis is found to be 4.35 $meV$ which is very small in comparison 
with the Mn binding energy in GaAs (113 $meV$). At first, the presence of a magnetic anisotropy 
is surprising since there is no difference between the [001] and [010] or [100] 
directions in a Zinc-Blende crystal. The observed anisotropy can be explained by the use 
of periodic boundary conditions on finite clusters used in this paper \cite{scm_MnGaAs_paper1_prb09}. 
The influence of other Mn atoms in the area may indeed introduce a small magnetic anisotropy 
and thus the observed magnetic anisotropy of Mn in bulk GaAs is artificial.
Indeed, more recent calculations carried on out on much larger clusters show that the bulk magnetic
anisotropy decreases monotonically with cluster size, down to a fraction of a meV for the
largest clusters of 40,000 atoms \cite{rm_cmc_2012}.

On the other hand, the calculation of the LDOS for Mn in bulk GaAs in Ref. \cite{scm_MnGaAs_paper1_prb09} shows good similarity 
with the calculations in Ref. \cite{tangflatte_prb05}. The LDOS is found to be spreading 
in the direction perpendicular to the Mn core spin axis. The change in the shape of 
the LDOS is explained in terms of the p$_{x}$, p$_{y}$ and p$_{z}$ character of the Mn acceptor hole. 
For different orientations of the Mn core spin, different components in the character dominate. 
When the Mn core spin direction is changed from [1$\bar{1}$0] to [110] a drop 
in LDOS of 74$\%$ is observed at 4 atomic layers from the Mn position. 
This drop in LDOS is 25$\%$ when the core spin direction changes from [1$\bar{1}$0] to [001], 
which is again in good agreement with the other calculations in \cite{tangflatte_prb05}.

In Ref.~\onlinecite{scm_MnGaAs_paper1_prb09} similar calculations have been done for 
Mn in or below the GaAs (110) surface layer. 
For Mn at the surface and the first subsurface layer, a strong localization of 
the LDOS is observed and a magnetic easy axis in the [111] direction is found. 
The difference in LDOS for different Mn core spin orientations is negligible. 
Thus in an X-STM experiment, we expect to see no effect of the magnetic field on 
the Mn atoms very close to the surface.

For Mn atoms deeper below the (110) surface, the LDOS becomes more extended and 
the magnetic anisotropy shows a complex behavior for subsequent depths. 
However, from the fourth layer beneath the (110) surface and deeper, one can recognize the 
emergence of an easy plane with its normal in the [1$\bar{1}$0] direction. 
The anisotropy energy is found to be at least 15 meV. Images of the (110) surface LDOS 
show that there is an increasing difference in LDOS for an increasing depth 
when the Mn core spin changes from the easy axis to the hard axis. 
For Mn atoms placed on fourth subsurface layers and deeper, the difference in LDOS varies between 40$\%$ and 82$\%$.

In summary, both Refs. ~\onlinecite{tangflatte_prb05} and ~\onlinecite{scm_MnGaAs_paper1_prb09} 
have treated the behavior of the Mn acceptor hole LDOS 
in the (110) plane for different Mn core spin orientations. In both papers it is found that 
when the Mn core spin direction is changed from [1$\bar{1}$0] to [110], a drastic change 
in the LDOS is taking place. 
The inclusion of the cleavage surface relaxation has resulted in similar observations.

The mechanism for the magnetic anisotropy in Refs.~\onlinecite{tangflatte_prb05} 
and \onlinecite{scm_MnGaAs_paper1_prb09} is the same --- the presence of the surface, 
or strain, lowers the energy of an orbital wave function with quantization axis along 
a specific direction, and the spin-orbit interaction (which correlates the spin axis
with the orbital axis) causes that preferred orbital direction to select a preferred spin axis. 
The effective energy associated with the correlation between spin axis and 
orbital axis is of the same order as the binding energy 
(Refs.~\onlinecite{FlattePRL2004, tangflatte_prb05,scm_MnGaAs_paper1_prb09} 
found it to be $\sim 40$~meV). 
At magnetic fields required to overcome the magnetic anisotropy energy 
the magnetic length is of order $3$~nm, which is three times larger than the 
effective Bohr radius of the acceptor ($\sim 1$~nm). Therefore the overall distortion 
of the acceptor state wave function due to the direct effect of the magnetic field 
on the orbital wave function is small compared with the spin-orbit term. 
What is not certain, however, is whether the effect of the magnetic field 
on the acceptor state wave function can substantially change the magnetic anisotropy; 
this will be examined in Section~\ref{theo_results}.

In an X-STM experiment, one 
can also check the results of
these calculations by applying a 
magnetic field perpendicular and parallel to the (110) cleavage plane and by measuring 
the Mn contrast, which can change with a factor as high as 90$\%$. 
In the next section, we discuss the X-STM experiments that have been performed 
to observe the predicted effects.
\section{Experiments}
\label{exp_results}
\begin{figure}
\includegraphics[height=130mm,width=80mm]{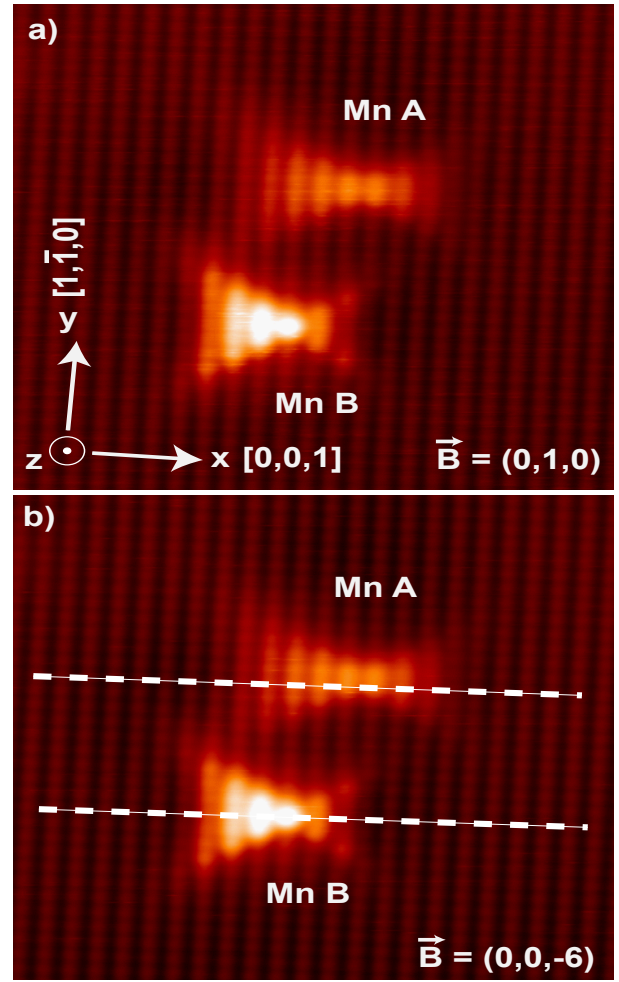}
\caption{13x13 nm X-STM images of two Mn atoms at different depths below the (110)
 cleavage plane. 
The x direction corresponds with the crystallographic [001] direction and the y-direction 
corresponds with the [1$\bar{1}$0] direction. The images are taken at +1.4 V and 50 pA 
at a temperature of about 2.5 K. In a) a magnetic field of 1 T is oriented 
in the [1$\bar{1}$0] y-direction and in b) a magnetic field of -6 T is oriented 
in the [110] z-direction.}
\label{Figure1}
\end{figure}

A Mn doped layer of 500 $nm$ thick was grown on a p-type GaAs substrate at high temperature with Mn concentrations of about 3$\times$10$^{19}$ $cm^{-3}$. The experiments are performed with an Omicron Cryogenic STM operating at a base temperature of 2.5 K. A magnetic field vector can be applied with fields of up to 6 T in the z-direction only or max. 2 T in the z-direction together with max. 1 T in the x- and y-directions. The x-, y- and z-direction are indicated in figure \ref{Figure1}a where 2 Mn atoms at different depths below the cleaved surface are visible.
The magnetic field is indicated in the vector notation in units of T: $\vec{B}$=($B_{x}$,$B_{y}$,$B_{z}$).\\
From Ref. \cite{celebi_prl10}, we estimate that Mn A is approximately 8 atomic layers 
below the cleavage surface and that Mn B is at about 5 atomic layers below the cleavage surface. 
In \cite{scm_MnGaAs_paper1_prb09}, a change in contrast of 40$\%$ is predicted for a 
Mn A at 8 layers below the cleavage surface when the Mn core spin changes from the 
[110] direction to the [1$\bar{1}$0] direction. For Mn B at 5 atomic layers beneath 
the cleavage surface, a change of 60$\%$ is predicted when the Mn core spin direction 
changes from the [110] direction to the [1$\bar{1}$0] direction. 
As can be seen from the comparison of figures \ref{Figure1}a and \ref{Figure1}b, 
there is no change at all in the Mn contrast for both Mn atoms when the magnetic 
field is changed from 1 T in the y direction to -6 T in the z-direction.
\begin{figure}
\begin{center}
\includegraphics[height=130mm,width=80mm]{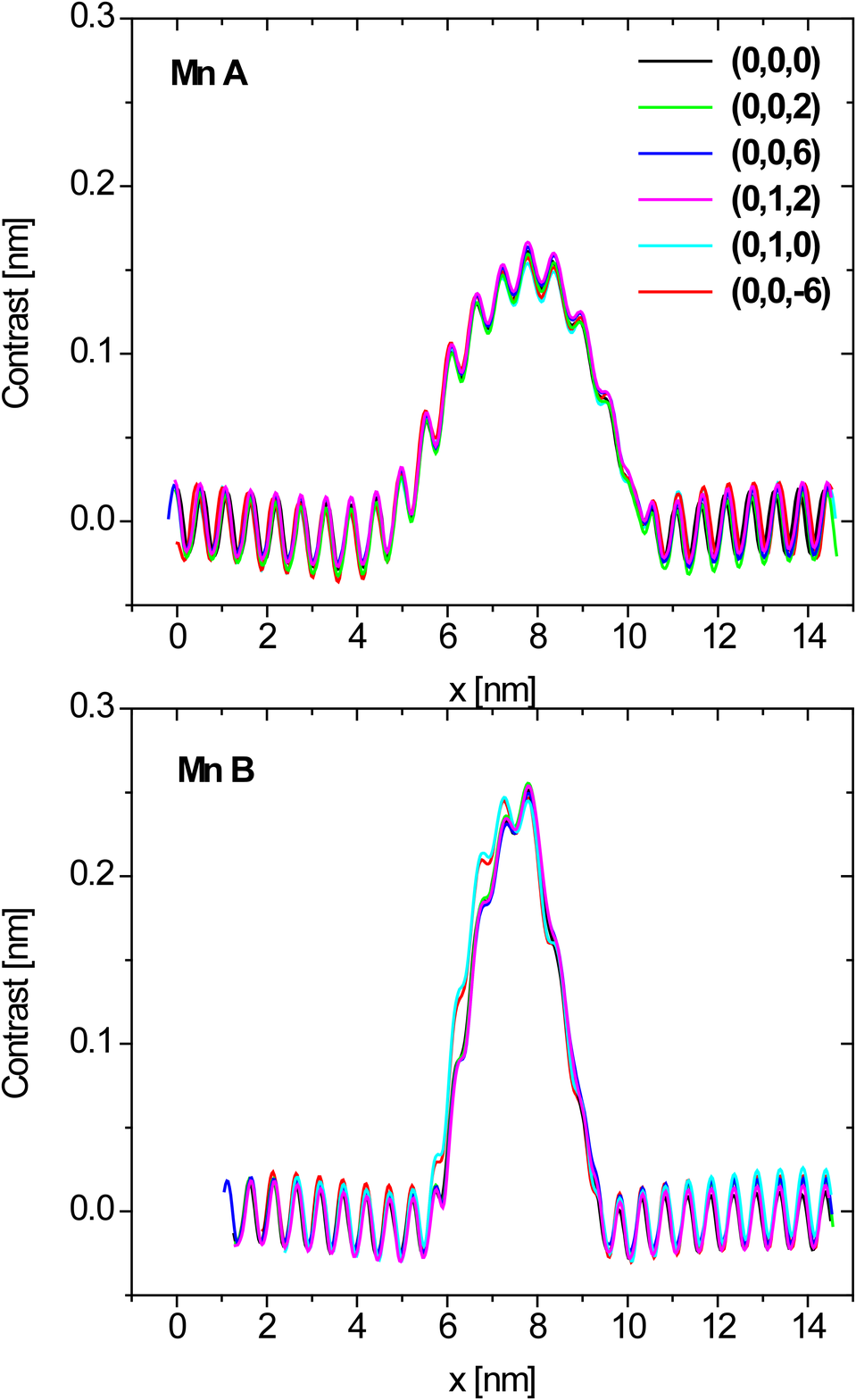}
\caption{a) Contrast of Mn A along the [001] direction indicated with a dashed line in figure \ref{Figure1}b. 
b) same plot for Mn B. Mn A has FWHM of about 4.5 $nm$ and Mn B has a FWHM of about 2.0 $nm$}
\label{ULCombined}
\end{center}
\end{figure}
In figure \ref{ULCombined}, a more quantitative comparison is made by looking 
at the contrast of the Mn atoms in different magnetic fields through the 
dashed lines in figure \ref{Figure1}b. Also in these plots, it can be seen that 
for both Mn atoms, there is no difference at all in the contrast for different orientations of the magnetic field.\\
For Mn B, the plots for B=(0,1,0) and B=(0,0,-6) are slightly different from the rest 
because of a small tip modification that has taken place. 
The tip has become slightly less sharp in the scan direction (the [001] x-direction) and 
this difference is noticed when sharper objects like Mn B are imaged. 
Mn A has FWHM of about 4.5 $nm$ in the scan direction,
while Mn B has a sharper feature with a FWHM of about 2.0 $nm$. The different FWHM of
the Mn features has been related to the depth below the GaAs surface  \cite{celebi_prl10} \\

\section{THEORETICAL MODEL}
\label{theo_model}
We model theoretically substitutional Mn impurities in GaAs following the procedure
put forward in Ref.~\onlinecite{scm_MnGaAs_paper1_prb09}.
Our second-quantized tight-binding Hamiltonian for
(Ga,Mn)As takes the following form:%
\begin{align}
H &  =\sum_{ij,\mu\mu^{\prime},\sigma}t_{\mu\mu^{\prime}}^{ij}a_{i\mu\sigma
}^{\dag}a_{j\mu^{\prime}\sigma}+J_{pd}\sum_{m}\sum_{n[m]}\vec{S}_{n}\cdot
\hat{\Omega}_{m}\nonumber\\
&  +\sum_{i,\mu\mu^{\prime},\sigma\sigma^{\prime}}\lambda_{i}\langle\mu
,\sigma|\vec{L}\cdot\vec{S}|\mu^{\prime},\sigma^{\prime}\rangle a_{i\mu\sigma
}^{\dag}a_{i\mu^{\prime}\sigma^{\prime}}\nonumber\\
&  +\frac{e^{2}}{4\pi\varepsilon_{0}\varepsilon_{r}}\sum_{m}\sum_{i\mu\sigma
}\frac{a_{i\mu\sigma}^{\dag}a_{i\mu\sigma}}{|\vec{r}_{i}\mathbf{-}\vec{R}%
_{m}|}+V_{\rm Corr},\label{hamiltonian}%
\end{align}
where $i$ and $j$ are atomic indices that run over all atoms, $m$ runs over
the Mn, and $n[m]$ over the nearest neighbors of Mn atom $m$. $\mu$ and $\nu$
are orbital indices and $\sigma$ is a spin index. The first term in
Eq.~(\ref{hamiltonian}) contains the near-neighbor Slater-Koster tight-binding
parameters\cite{slaterkoster_pr54,papac_jpcm03} that reproduce the
band structure of bulk GaAs\cite{chadi_prb77} and that are
rescaled\cite{chadi_prl78,chadi_prb79,scm_MnGaAs_paper1_prb09} when needed to
account for the buckling of the (110) surface. 

The second term implements the
antiferromagnetic exchange coupling between 
the Mn spin $\hat{\Omega}_{m}$ (treated as
a classical vector) and the nearest neighbor 
As $p$-spins $\vec{S}_n =  1/2\sum_{\pi\sigma{\sigma'}} a_{n\pi\sigma}^\dagger 
\vec{\tau}_{\sigma{\sigma'}} a_{n\pi{\sigma'}}$.
The exchange coupling $J_{pd}=1.5$ eV has been inferred from
theory \cite{timmacd_prb05} and experiment \cite{ohno_sci98}. As a result of
this term the acceptor hole that is weakly bound to the Mn will become spin
polarized. 
This
model contains only $s$ and $p$ orbitals, and the effect of the Mn $3d^{5}$
electrons is encoded in the exchange term.

Next, we include an on-site spin-orbit one-body term, where the
renormalized spin-orbit splittings are taken from
Ref.~\onlinecite{chadi_prb77}. Spin-orbit coupling will cause the total
energy to depend on the Mn spin direction, defined by a collinear variation of
$\hat{\Omega}_{m}$. 

The fourth term is a long-range repulsive Coulomb part
that is dielectrically screened by the host material. To account in a simple way for weaker
dielectric screening at the surface, the dielectric constant $\epsilon_r$ for a Mn on the
surface is reduced from the bulk GaAs value 12 to 6 for the
affected surface atoms. This crude choice is qualitatively supported by
experimental results \cite{Teichmann_prl08,Gupta_NanoLett}.\ 

The last term $V_{\rm Corr}$ is 
a one-particle correction potential
for the Mn central cell. This term is the least known and understood theoretically. It consists 
of on- and off-site
parts, $V_{\mathrm{\rm corr}}=V_{\mathrm{\rm on}}+V_{\mathrm{\rm off}}$ 
which influence
the Mn ion and its As nearest neighbors respectively. The on-site Coulomb
correction is estimated to be $1.0$ eV from the ionization energy of Mn. The
off-site Coulomb correction affects all the nearest-neighbor As atoms surrounding
the Mn ion and together with the exchange interaction, it reflects
primarily the $p$-$d$ hybridization physics and is the parameter 
that in the model primarily controls the binding energy of the
hole acceptor state.
The off-site Coulomb correction value is set by tuning the position of the
Mn-induced acceptor level in the bulk to the experimentally observed
position\cite{schairer_prb74,lee_ssc64,chapman_prl67,linnarsson_prb97} at 113
meV above the first valence band level. 
The value thus obtained is
$V_{\mathrm{\rm off}}=2.4$ eV. 
When the Mn impurity is on the GaAs surface, the value
of $V_{\mathrm{\rm off}}$ is reduced to ensure that the position of the acceptor
level is consistent with the value attained via STM spectroscopy. 

The off-site Coulomb correction is in fact a repulsive potential for the electrons. 
If we use the bulk value (2.4 eV) for the surface, the acceptor level lies deep in 
the gap at 1.3 eV above the valence band,  which means the acceptor wave function 
is now much more localized around the Mn than its bulk counterpart. In order to 
guarantee the experimentally observed position for the acceptor level, 0.85 eV 
\cite{yazdani_nat06}, we have to decrease this repulsive potential for the 
electrons, which causes the hole wave function to be less localized with a corresponding smaller
binding energy.

The electronic structure of GaAs with a single substitutional Mn atom 
is obtained by
performing a super-cell type calculation with a cubic cluster of a  few thousands atoms
and periodic boundary conditions in either 2 or 3 dimensions, depending on
whether we are studying the $\left(  110\right)  $ surface or a bulk-like
system. The $\left(  110\right)  $ surface of GaAs is simplified from both
theoretical and experimental points of view, by the absence of large surface
reconstruction. In order to remove
artificial dangling-bond states that would otherwise appear in the
band gap, we include relaxation of surface layer positions 
following a procedure introduced in 
Refs.~[\onlinecite{chadi_prl78, chadi_prb79}].
For more details the reader is
referred to Ref.~[\onlinecite{scm_MnGaAs_paper1_prb09}].

We would like to emphasize that the strength of the off-site Coulomb correction is the 
only important fitting parameter of the model,
and its value is fixed once for all by the procedure described above. 
All the other parameters in Eq.~\ref{hamiltonian} are either determined
by theoretical considerations, or for the cases when this is not possible 
(e.g. short- range onsite potential) their values
are extracted from experiment. In any case, they affect weakly the properties of the acceptor level.
Once the parameters of the Hamiltonian of
Eq.~\ref{hamiltonian}
are chosen in the way indicated above, the model has to be viewed as a microscopic description,
with predictive power, of the properties of Mn impurities in GaAs surfaces and subsurfaces.
In this sense the model of Ref.~\onlinecite{scm_MnGaAs_paper1_prb09} has been quite successful 
in capturing some of the salient features
of the STM experiments \cite{koenraad_prb08, GarleffPRB2010}, probing the Mn-dopant acceptor hole 
near the GaAs (110) surface. For example, 
it correctly 
describes the dependence of the acceptor binding energy \cite{GarleffPRB2010} 
and the shape of the hole wave function \cite{koenraad_prb08} on the layer depth below 
the surface on which the magnetic dopant is positioned. The model also makes a prediction
on how the magnetic anisotropy barrier for the Mn-impurity--hole magnetic complex changes as a function
of the layer depth. These predictions can be indirectly checked by the magnetic-field studies that are
the main scope of the present paper.

In order to study the response 
of the system to an external magnetic field, we introduce the Zeeman term
\begin{eqnarray}
H_{z} &=& - \frac{\mu_B}{ \hbar} \sum_i \sum_{\mu{\mu'}\sigma{\sigma'}} 
 \bigg\langle\mu\sigma
\bigg|(\vec L+g_s\vec S)\cdot \vec B\bigg|{\mu'}{\sigma'}
\bigg\rangle  a_{i\mu\sigma}^\dagger a_{i{\mu'}{\sigma'}} \nonumber \\
 && -g_s \frac{\mu_B}{\hbar} \sum_m  \hat{\Omega}_m \cdot \vec B\;,
\label{eq:six} 
\end{eqnarray}
where the first term runs over all $s$ and $p$ orbitals of all atoms, 
and the second term represents the coupling of the magnetic field with the magnetic moment of the Mn 
impurities, treated as a classical vector. 
Here $\mu_B=  \frac{\hbar e} {2m} = 5.788 \times 10^{-2}\ {\rm meV\ T}^{-1}$ is the Bohr magneton, $g_s = 2$,
and we follow the incorrect but common convention that spins and magnetic moments are parallel to each other
\footnote{Because the electron charge is negative, 
magnetic moments and angular momentum are in fact oriented antiparallel to each other. 
In a magnetic field the energetically favorable direction of the magnetic moment is parallel to the field
while the direction of the spin is antiparallel. Assuming that magnetic moment and angular momentum are
parallel is strictly speaking incorrect but does not change the physics.}.
Therefore in the paper we will loosely refer to the direction of $\hat \Omega$ as the direction
of the Mn magnetic moment.

\section{THEORETICAL RESULTS AND DISCUSSION}
\label{theo_results}
We start by analyzing the magnetic anisotropy properties for one Mn at the (110) GaAs surface layer and the immediate
subsurface layers, and see how these are modified by the presence of an external magnetic
field of a few Tesla. The magnetic anisotropy landscape as a function of $\hat \Omega$ 
for one Mn at the surface and the first 9 subsurfaces
has been studied in detail in Ref.~\onlinecite{scm_MnGaAs_paper1_prb09}. Typically the system has an uniaxial
anisotropy  with two minima separated by an energy barrier. We will refer to the $\hat \Omega$ direction of
minimum energy as the {\it easy} direction and the one of maximum energy as the {\it hard} direction. 

We first consider the case of one Mn impurity at the (110) surface.
To facilitate the comparison with the case in which a magnetic field is present, we recalculated
and plotted here anisotropy landscapes and LDOS in the absence of the magnetic field, originally published 
in Ref.~\onlinecite{scm_MnGaAs_paper1_prb09}, using an improved code.

\begin{figure}[h]
\centering
\includegraphics[scale=0.17]{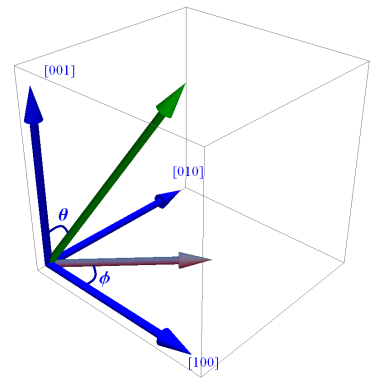}
\caption{Color online -- The direction of $\theta$ and $\phi$ with respect to the crystal axis.}
\label{fig:crystal}
\end{figure}

\begin{figure}
\centering
\includegraphics[scale=0.16]{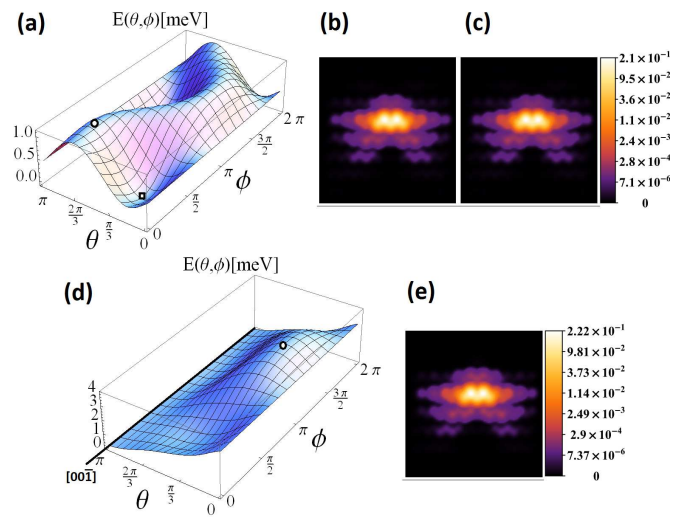}
\caption{Color online -- The magnetic anisotropy energy (MAE) landscape and the Mn-acceptor-level 
local density of states (LDOS) for one Mn at 
the [110] surface. 
(a) The MAE in the absence of an external magnetic field, as a function
of the angles $\theta$ and $\phi$ defining the direction of the Mn spin $\hat \Omega$. The barrier (hard)
direction is marked with a circle and the minimum energy (easy direction) with a square.
(b) and (c) LDOS of the Mn acceptor level when the Mn magnetic moment 
points in the easy and hard direction respectively, as defined in panel (a). (d) The MAE in the presence 
of a 6 T external magnetic field applied along the hard direction ($ \theta=3\pi/4,\phi=\pi/4 $). 
(e) The LDOS
in the presence of a 6 T magnetic field. Here the Mn magnetic moment is along the easy direction determined by
the landscape (d) modified by the presence of the external field. The barrier (hard)
direction is marked with a circle and the minimum energy (easy direction) with a solid line.}
\label{fig:Surface}
\end{figure}
In Fig.~\ref{fig:Surface}(a) we plot the anisotropy energy landscape in the absence of the magnetic field, as a function
of the angles $\theta$ and $\phi$ defining the direction of $\hat \Omega$.
The coordinate system used for this and the other plots in the paper has $\theta=0$ parallel to the [001] axis,
$\left(  \theta=\pi/2,\phi=0\right)  $ parallel to [100], and
$\left( \theta=\pi/2,\phi=\pi/2\right) $ parallel to [010]. 
See Fig.~\ref{fig:crystal}.

The anisotropy landscape displays two minima, identifying the easy direction [111], separated by an energy barrier
of the order of 1 meV. Note that these tight-binding results of the
magnetic anisotropy of a Mn at the (110) GaAs surface are consistent with recent 
first-principles estimates\cite{fi_cmc_prb_2012}. 
Panels (b) and (c) of Fig.~\ref{fig:Surface} show the LDOS for the Mn acceptor state
when the Mn spins point along the easy and hard direction respectively, determined from the landscape in (a).
As discussed in Sec. II and shown clearly in the figures, 
the acceptor state wavefunction for a Mn on the surface is very localized around the impurity,
and the dependence of the LDOS on the Mn spin orientation is negligible.
The acceptor wavefunction, itself strongly anisotropic, seems to be completely decoupled from the
orientation of the Mn magnetic moment. 

Fig.~\ref{fig:Surface}(d) and (e) show the effect of a 6 T magnetic field on the anisotropy and LDOS
respectively, when the field is 
applied in the hard direction of the anisotropy landscape in (a).

We can see that the magnetic field changes considerably the anisotropy landscape, which has now an easy axis
at $\theta = \pi$ (the direction of the field). Note that in the presence of the field the anisotropy barrier 
has increased up to $~ 4$ meV.
The LDOS in Fig.~\ref{fig:Surface}(e) is now calculated for $\hat \Omega$ pointing along
the new easy axis, determined by the magnetic field. Despite the strong change in the anisotropy landscape
brought about by the magnetic field, the acceptor LDOS is essentially identical to the one calculated in the
absence of the field, in agreement with the experimental results.

\begin{figure}
\centering
\includegraphics[scale=0.23]{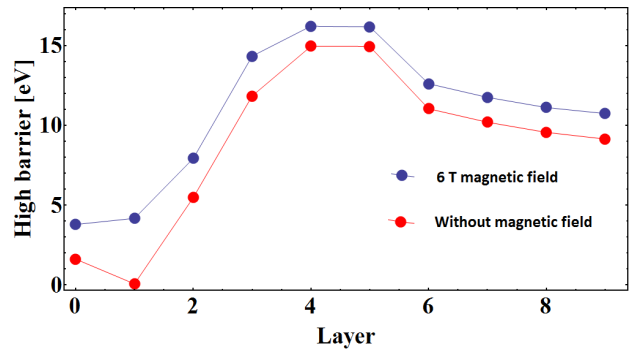}
\caption{Color online -- The maximum MAE barrier height
as a function of the Mn depth. Red dots are the MAE barrier 
height in the absence of an external magnetic field, while blue dots represent the height in the 
presence of a 6 T external magnetic field.}
\label{fig: MAE vs layer}
\end{figure}

Before continuing our LDOS analysis, it is useful to consider how
the anisotropy-energy barrier depends on the Mn-impurity depth from the (110) surface.  
In Fig.~\ref{fig: MAE vs layer} we plot the largest value of the anisotropy energy barrier as a function
of the subsurface layer index (layer 0 is the (110) surface). In general, 
in the absence of a magnetic field (red dots in the picture) the anisotropy barrier increases
with Mn depth, reaching a maximum of $~15$ meV for layers 4 and 5. It then starts to decrease and it should eventually 
reach a very small value corresponding to the case where the Mn is effectively in the bulk.
For the finite clusters that we have considering here (20 layers in the z-direction), the
anisotropy remains large also when the impurity is effectively in the middle 
of the cluster (corresponding to layer 9 from the surface). Bulk calculations on considerably larger clusters show that the anisotropy for impurities in the middle of the clusters does decrease to a fraction
of one meV\cite{rm_cmc_2012}. 
For these larger clusters the magnetic anisotropy of the Mn positioned in
on layers $\ge 8$ is expected to decrease a bit with cluster size. However the qualitative
behavior of the first 7-8 layers shown in Fig.~\ref{fig: MAE vs layer}, and the corresponding numerical
values of the magnetic anisotropy are controlled by the vicinity to the surface and as such
should not depend strongly on cluster size\cite{rm_cmc_2012}.

\begin{figure}[h]
\centering
\includegraphics[scale=0.8]{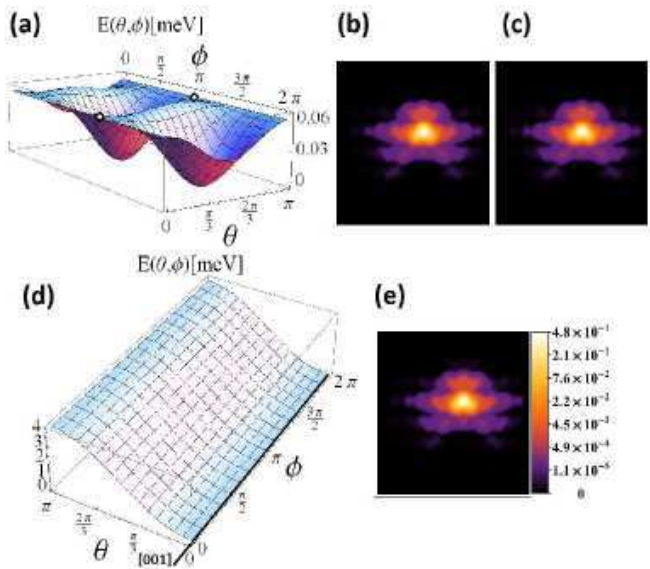}
\caption{The MAE landscape and the Mn-acceptor-level LDOS for one Mn in the 
first subsurface (i.e., one layer below the [110] surface). (a) The MAE in the 
absence of an external magnetic field. The barriers (hard)
directions are marked with a circle. (b) and (c) the Mn acceptor LDOS
for the case in which the Mn magnetic moment points in the easy and hard 
direction respectively. (d) The MAE in the presence of a 6 T external 
magnetic field pointing along the (original) hard direction ($ \theta=0,\phi=\pi $).
The minimum energy (easy direction) is shown with a solid line. 
(e) The Mn acceptor LDOS
in the presence of a 6 T magnetic field. 
Here the Mn magnetic moment points in the new easy direction
determined by the magnetic field, as shown in (d). The colorscale in (b) and (c) is the same as in (e).}
\label{fig:First sublayer}
\end{figure}

Layer 1 (the first subsurface layer)is a special case
in the sense that the anisotropy is very small, on the order of 0.1 meV. 
The first subsurface represents the cross over from the case in which the Mn is at the surface, with 
three nearest neighbor As, to a bulk-like
environment characterized by four nearest neighbor As atoms. The properties of the acceptor level 
found in STM experiments for a Mn positioned on this 
subsurface are also quite anomalous\cite{gupta_science_2010}. 
When a magnetic field of 6 T is applied along the hard direction
(blue dots in Fig.~\ref{fig: MAE vs layer}) the anisotropy barrier increases by a couple of meV. 
The exception is again the first subsurface (layer 1), whose anisotropy is now 
completely controlled by the magnetic field and behaves in a similar way to the surface layer.
The behavior of the first subsurface anisotropy landscape is shown explicitly in Fig.~\ref{fig:First sublayer} (a), (d). 

As for the case of a Mn atom placed at the (110) surface, 
the acceptor LDOS for a Mn on the first subsurface (see Fig.~\ref{fig:First sublayer}(b), (c) ) 
is completely insensitive to the direction of the Mn magnetic moment. Again a 6 T magnetic field, which is able to
completely modify the magnetic anisotropy landscape and orient the Mn moment parallel to its direction, does not have
any detectable effect on the acceptor wave function, as shown in Fig.~\ref{fig:First sublayer}(d).

\begin{figure}
\centering
\includegraphics[scale=0.8]{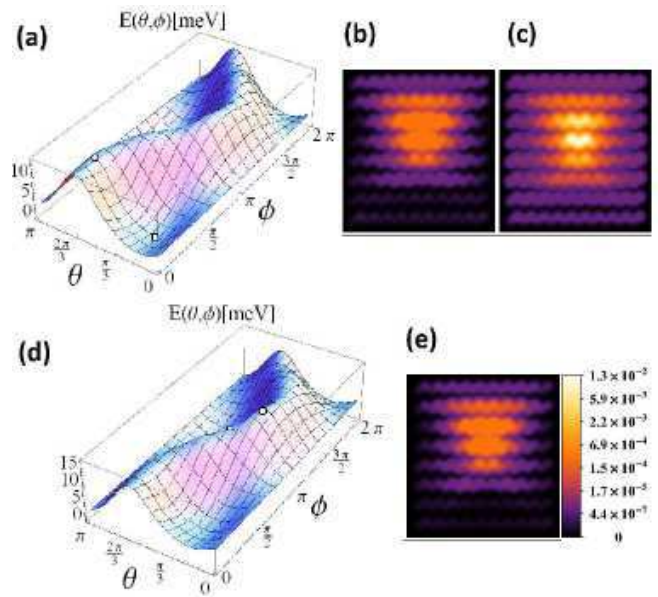}
\caption{The magnetic anisotropy energy (MAE) and the LDOS for one Mn in the 
fourth subsurface (fourth layer below the [110] surface). (a) The MAE in the 
absence of an external magnetic field. (b) and (c) the LDOS of Mn acceptor 
level for the case that Mn magnetic moment points in the easy and hard direction respectively. 
(d) The MAE in the presence of a 6 T external magnetic field which points along the hard direction($ \theta=\pi/2,\phi=3\pi/4 $). 
(e) The LDOS in the case that magnetic moment points in the easy direction in the presence of a 6 T magnetic field.
The barrier direction in (a) and (d) is marked with a circle and the easy direction with a square. The colorscale in (b) and (c) is the same as in (e).}
\label{fig:Fourth sublayer}
\end{figure}

\begin{figure}[h]
\centering
\includegraphics[scale=0.8]{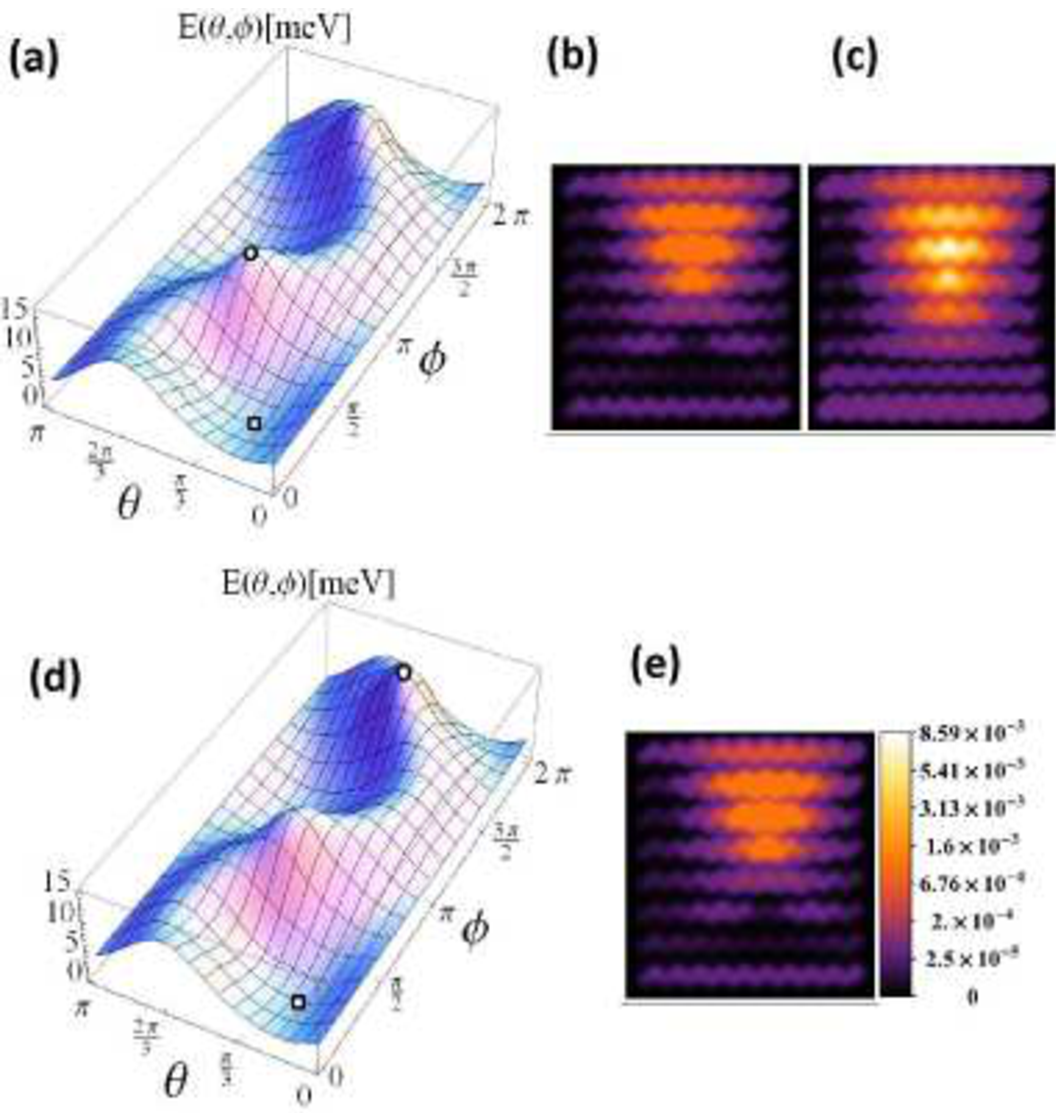}
\caption{As in Fig.~\ref{fig:Fourth sublayer}, but for the fifth subsurface (fifth layer below the
 [110] surface). The colorscale in (b) and (c) is the same as in (e).}  
\label{fig:Fifth sublayer}
\end{figure}
As the Mn is placed in successively deeper layers below the surface and the acceptor wavefunction becomes less localized 
around the impurity, the situation changes. In Figs.~\ref{fig:Fourth sublayer} and \ref{fig:Fifth sublayer}
we plot the anisotropy landscape and the acceptor LDOS for the fourth and fifth subsurface 
(layer 4 and 5 below the surface)
respectively. As we discussed before, when the direction of the Mn moment is forced to point in the
hard direction (panel (c) of Fig.~\ref{fig:Fourth sublayer}) the LDOS around the Mn increases sensibly. 
The two cases, easy (panel b) and hard axis LDOS are now clearly distinguishable. Since the acceptor wavefunction
is always normalized, an increase of the LDOS in the core region implies that the acceptor 
wavefunction is considerably
more localized
when the Mn magnetic moment points in the hard direction. 
On the other hand, in contrast to the surface and the first subsurface, the energy barrier in this case is considerably
larger. A magnetic field of the order of those applied experimentally are now not strong enough to modify appreciably
the anisotropy landscape. This can be seen
by comparing panel (a) -- no magnetic field -- with panel (d), where magnetic field of 6 T 
is applied in the hard direction.
Consequently, the direction of the easy axis is only slightly modified in the presence of a magnetic field, and as a result, the corresponding acceptor LDOS appears now very similar to the zero-magnetic field case [panel (b)].
This is again in agreement with the experiments presented in this paper.

As mentioned before, the experiments presented in \cite{GarleffPRB2010} showed the energy level splitting of Mn in GaAs close to the cleavage surface. For a typical Mn position at 5th subsurface layer, a total splitting of 14 meV is found between the 3 peaks which are attributed to the different projections of the total momentum J=1 which is the result of anti-ferromagnetic coupling between the 5/2 Mn core spin and 3/2 Mn acceptor total angular momentum. In Fig. \ref{fig:Fifth sublayer}, it can be seen that the MAE is indeed about 15 meV which corresponds well with the findings in \cite{GarleffPRB2010}.

In Fig. \ref{fig:JMing}, the MAE as calculated in another tight binding calculation (strained bulk GaAs) is about 23 meV. This is more than the 15 meV of the supercell calculations (Fig. \ref{fig:Fifth sublayer}) possibly because of the overestimated strain or its assumed uniformity.

We conclude that, although the LDOS of deep-subsurface Mn acceptors is in principle 
strongly dependent on the Mn magnetic moment direction, its actual manipulation with an external magnetic field
is not suitable at field strengths presently used in experiment.

\section{Conclusions}

In conclusion, this work is the first systematic study of the effect of an external 
applied magnetic field on the acceptor properties of individual Mn impurities in GaAs. 
Specifically, we have investigated theoretically and experimentally 
the effect of an external magnetic field on the acceptor hole wavefunction and LDOS
of Mn impurities placed near the (110) surface of GaAs. The acceptor LDOS is directly accessible via
X-STM experiments. 

The motivation of this study was in part provided by previous theoretical studies which predicted that
the LDOS in some cases strongly depends on the orientation of the magnetic impurity magnetic moment. 
The theoretically model used in this analysis is essentially parameter-free, 
once the energy of the surface acceptor state is fixed to reproduce the experimental value.

Experimentally we find that there is no detectable difference in the STM images of the acceptor hole LDOS when
a magnetic field up to 6 T is applied in several directions with respect to the crystal structure.
To reconcile theory and experiment we have carried out a theoretical analysis of the magnetic anisotropy energy
and acceptor hole wavefunction in the presence of a magnetic field. We have shown that for Mn impurities placed in 
deep sub-layers below the surface, the calculated magnetic anisotropy landscape is characterized by energy barriers
of the order of 10-20 meV, which are only minimally affected by magnetic fields used in experiment.
We estimate that one needs to employ much stronger fields 
(on the order of tens of Tesla) to modify significantly the 
anisotropy landscape and rotate the magnetic moment of the impurity. This estimate is based on the idea of manipulating a spin=5/2 object with g-factor=2 with an external field to overcome an energy barrier of 15 meV.

For impurities placed near the surface,
the magnetic anisotropy is small enough to be considerably affected by a magnetic field of a few T. However,
for this case the acceptor hole LDOS is much less sensitive to the orientation of the Mn magnetic moment.
The combination of these two facts seem to explain the experimental finding that the the STM images of the
acceptor hole wavefunction is essentially unaffected by an external magnetic field.

Our studies show that the Mn-dopant behavior close to the GaAs surface depends on the
layer depth in a complex and highly non trivial way. These studies also suggest that it could be 
interesting to carry out a similar investigation
for other magnetic dopants and other semiconductors. 
It might be possible that for 
some of these systems the acceptor wavefunction for a dopant near the surface 
be more delocalized and amenable to an easier manipulation 
by a static magnetic field, displaying the effects 
originally predicted for Mn in GaAs. It should also be possible 
to use resonant techniques, such as those commonly used 
in electron spin resonance and ferromagnetic resonance, 
to map out the anisotropy landscape presented 
here for Mn near the GaAs surface. 
Finally, excitations of the spin that would correspond 
to the quantized spin in the anisotropy landscape 
here should be visible in inelastic tunneling spectroscopy. 
Thus these new predictions do not mean that Mn spin dynamics 
is impossible to see near the surface of GaAs, 
merely that it is more challenging to observe.

\begin{acknowledgments}
We would like to thank W. van Roy and Z. Li for the Mn doped GaAs samples which they have provided and A. H. MacDonald for several useful discussions.
We would like to thank also B. Bryant for substantial experimental support during the X-STM measurements.

This work was supported by STW-VICI Grant No. 6631. We acknowledge support from the Engineering and Physical Sciences Research Council (EPSRC) through grants EP/D063604/1 (PS, SRS, NJC, and CFH) and EP/H003991/1 (SRS). This work was also supported  in part by the Faculty of Natural Sciences at Linnaeus University,
by the Swedish Research Council under Grant Numbers: 621-2007-5019
and 621-2010-3761 and by the Nordforsk research network: 08134, {\it
Nanospintronics: theory and simulations}.
\end{acknowledgments}

\bibliography{mnbiblio_Murat}

\end{document}